\documentstyle[12pt]{article}
\topmargin - 0.5in
\oddsidemargin - 5mm
\begin{document}

\def\p{\phi}
\def\P{\Phi}
\def\a{\alpha}
\def\e{\varepsilon}
\def\be{\begin{equation}}
\def\ee{\end{equation}}
\def\l{\label}
\def\0{\setcounter{equation}{0}}
\def\T{\hat{T}_}
\def\b{\beta}
\def\S{\Sigma}
\def\C{\cite}
\def\r{\ref}
\def\ba{\begin{eqnarray}}
\def\ea{\end{eqnarray}}
\def\n{\nonumber}
\def\R{\rho}
\def\q{\hat{Q}_0}
\def\X{\Xi}
\def\x{\xi}
\def\la{\lambda}
\def\d{\delta}
\def\s{\sigma}
\def\f{\frac}
\def\D{\Delta}
\def\pa{\partial}
\def\Th{\Theta}
\def\o{\omega}
\def\O{\Omega}
\def\th{\theta}
\def\ga{\gamma}
\def\Ga{\Gamma}
\large
\begin{titlepage}
\vskip 1cm
\begin{center}
{\Large\bf Wigner functions of essentially nonequilibrium systems}
\vskip 1cm

J.Manjavidze\\ Institute of Physics, Georgian Academy of Sciences, \\
Tamarashvili st. 6, Tbilisi 380077, Republic of Georgia, \\
e-mail:~jm@physics.iberiapac.ge
\end{center}

\begin{abstract}

The aim of the article is to discuss the $S$-matrix interpretation of
perturbation theory for the Wigner functions generating functional at
a finite temperature. For sake of definiteness, fruitful from pedagogical
point of view, the concrete problem from
particle physics of high-temperature initial states dissipation into cold
one is considered from experimental and theoretical points of view.
The temperature is introduced in the theory by typical for the
microcanonical description way. The perturbation theory contains two-
temperature (of initial and final states) Green functions. Two possible
boundary conditions are considered. One of them is usual in a field theory
vacuum boundary condition. Corresponding generating functional of Wigner
functions can be used in the particle physics. Another type of the boundary
condition assumes that the system under consideration is in environment of
the black-body radiation. This leads to the usual in statistics
Kubo-Martin-Schwinger boundary condition at the equilibrium (one-temperature)
limit. The comparison of the $S$-matrix approach with
Schwinger-Keldysh real-time finite-temperature field theory and with
nonstationary statistical operator approach of Zubarev are considered.
The range of applicability of the finite-temperature description of
dissipation processes is shown.

\end{abstract}   \end{titlepage}

\section{Introduction}\0

At the very beginning of this century couple P. and T.Ehrenfest had offered
a model to visualize Boltzman's interpretation of irreversibility
phenomena in statistics. The model is extremely simple and fruitful \C{kac}.
It considers two boxes with $2N$ numerated balls. Choosing number
$l=1,2,...,2N$ $randomly$ one must take the ball with label $l$ from one
box and put it to another one. Starting from the highly `nonequilibrium'
state with all balls in one box it is seen tendency to equalization of
balls number in the boxes. So, there is seen irreversible\footnote{`What
never? No never! What never? Well, hardly
ever.' \C{May}} flow toward preferable (equilibrium) state. One can
hope \C{kac} that this model reflects a physical reality of nonequilibrium
processes with initial state very far from equilibrium. A theory of such
processes with (nonequilibrium) flow toward a state with maximal entropy
should be sufficiently simple to give definite theoretical predictions.

In order to do the consideration less formal we will be connected with
concrete physical problem. For instance,
the particles creation processes are good laboratory for investigation of
relativistic nonequilibrium processes general properties. Indeed, considering
the multiplicity $n$ as the characteristics of final state entropy we can
choose the asymptotically large $n>>\bar{n}(s)$, where mean multiplicity
$\bar{n}(s)$ naturally defines the scale of $n$. Then one can expect, noting
above mentioned general property of the nonequilibrium flow, that the theory
of processes with practically total dissipation of initial-state kinetic
energy into particles masses should be extremely simple. By this reason it
is natural to start consideration from region $n>>\bar{n}(s)$. We would
construct the theory permanently taking into account just this condition.

The theory of dissipative processes have general significance from
thermodynamical point of view and we would concentrate attention on this
important problem. There is also practical side of considered problem. At
$n>>\bar{n}(s)$ the cross sections $\s_n (s)$ falls down rapidly and are too
small ($< nb$). Noting also a problem of triggering such rear final state the
experimenters must have enough arguments to examine them. The main arguments
are as follows: at $n>>\bar{n}(s)$ we have unique chance (i) to examine \\
-- $pure$ (practically without admixture of hadrons),\\
-- $cold$ (it is a best condition for investigation of collective phenomena
in a system),\\
-- $dense$ (in this case the QCD interaction constant $\a_s$ is small) \\
quarks plasma (CQGP) and (ii) realize experimentally decay of very hot
(at high energies) initial state in the `inflational' regime, with `freezed'
nonperturbative degrees of freedom of hadrons system.

It is known from hadrons high-energy experiments that the cross sections
$\s_n$ have a maximum at $n \sim \bar{n}(s)$, where $1<< \bar{n}(s) <<n_{max}$
and $n_{max}=\sqrt{s}/m_h$ is the maximally available multiplicity at given
energy $\sqrt{s}$ ($m_h$ is the hadrons characteristic mass). This testify to
the statement that in hadron processes the nonequilibrium flow is not equal
to zero ($\bar{n}(s)>>1$), but the mostly probable processes did not lead to
the state with maximal entropy ($\bar{n}(s) <<n_{max}$). (The early models
was based on the assumption that the final state of inelastic hadron
processes has maximal entropy $\bar{n}(s) \sim n_{max}$ \C{fer}.)

The preferable at $n \sim \bar{n}$ processes are indebted for excitation of
hadrons nonperturbative degrees of freedom described by creation of hadrons
constituents from vacuum: the kinetic motion of partons leads to increasing,
because of confinement phenomenon, polarization of vacuum and to its
instability concerning quarks creation \C{str}. In other words, there is a
long-range correlation among hadrons constituents. Under this special
correlations the conservation laws constraints was implied. They are
important in dynamics since each conservation law decrease number of the
dynamical degrees of freedom at least on one unite, i.e. it has
nonperturbative effect (this must explain why $\bar{n}(s) << n_{max}$).
Moreover, in so-called integrable systems each independent integral of
motion (in involution) reduces number of degrees of freedom on two units.
In result there is not stochastization in such systems \C{zak}, i.e. the
nonequilibrium flow is equal to zero. But it will be argued that at the
very high multiplicities this effect is negligible. So, if
$$
\bar{n}(s) <<n<n_{max}
$$
we will see that the particles creation processes are close to Markovian
in accordance with Boltzman's idea. The reason of this phenomena is the more
fast falling down of soft chanel of hadrons creation comparing with hard
channels in asymptotics over $n$.

Rejecting nonperturbative effects creation of the high-multiplicity final
state can be described using standard methods of QCD. We will show dominance
of processes with minimal number of QCD jets in the high-multiplicities
region. This means that the high-multiplicity processes are stationary
Markovian\footnote{The vertexes renormalization takes into account the
time-reversed fluctuations in the nonequilibrium flow.}. This result is in
agreement with Boltzman's general idea concerning nonequilibrium flows.

So, the high-multiplicity processes are `unshadowed' by nonperturbative and
complicated perturbation effects. This will allow to investigate not only
new state of the pure colored plasma but also the structure of fundamental
Lagrangian. This conclusions are not evident and we start consideration from
brief review of arguments.

It must be noted that the experimental investigation of high-multiplicity
processes in deep asymptotics over $n$ seems unreal. But considering
moderate $n>>\bar{n}$ we can not be sure that the final state is
equilibrium. Investigation of fractal dimensions in the multiparticle
hadron processes at high energies shows presence of considerable
fluctuations \C{drem}. This leads to necessity to have the theory of
dissipation processes with nonequilibrium final state.

There is also another side of the problem. Today understanding of hadron
processes is far from ability to give any quantitative prediction. The
above announced prediction concerning absence of nonperturbative
contributions into hadron processes `work' in the deep asymptotics over
$n$ only. So, at moderate $n>>\bar{n}$ we can not be sure that they do
not have important influence. That is why we will concentrate our attention
in this paper on the searching of economic (thermodynamical) description of
the dissipative processes, trying to find the connections of our $S$-matrix
approach with other ones. It is important to note that the offered
formalism allows to separate the dynamical side of question from the pure
descriptional one (see also concluding Section).

This central problem of formalism can be solved noting that our dissipative
problem contains element of dynamics since it crucially depends from
boundary condition. Therefore, we adopt $S$-matrix formalism which is
natural for dissipative systems time evolution description. For this purpose
the amplitudes
$$
<(p)_m|(q)_n>=a_{n,m}(p_1,p_2,...,p_m;q_1,q_2,...,q_n)
$$
of the $m$- into $n$-particles transition will be introduced. (The
in- and out-states must be composed from mass-shell particles \C {pei}.)
Moreover, to incorporate the boundary condition $n>>m$ we should calculate
the probability integrating over particles momenta:
$$
r(P;n.m) \sim \int |a_{n,m}|^2= \int <(p)_m |(q)_n ><(q)_n |(p)_m >
$$
since the amplitude $a_{nm}(p_1,p_2,...,p_m;q_1,q_2,...,q_n)$ is the
function of too many variables. This standard method of particles physics
practically solves our problem.

Nevertheless it is desirable to use the thermodynamical language as the
mostly economic one, i.e. the formalism which uses minimal number of
parameters
(temperature, chemical potential, etc.) for description of the system.

The field-theoretical description of statistical systems at a finite
temperature is usually based on the formal analogy between imaginary
time and inverse  temperature $\b$ ($\b=1/T$) \C {bl}. This approach
is fruitful \C {mats} for description of the static properties of a system,
but it demands a complicated mathematical apparatus for the
analytic continuation to the real time \C {land}, if we want
to clear up dynamical aspects.
The  first important quantitative attempt to build the real-time
finite-temperature field theory \C {jack} discover a problem of the
pinch-singularities. Further investigation of the theory have
allowed to demonstrate the  cancellation mechanism of  these unphysical
singularities \C {sem}. This attained by doubling of the degrees of
freedom \C {sch, kel}:  the Green functions of the theory represent $2$
$\times$ $2$ matrix.  It surely makes the theory more complicated, but
the operator formalism of the  thermo-field dynamics \C {um} shows the
unavoidable character  of this complication.

The Schwinger-Keldysh real-time finite-temperature field-theoretical
de\-scription \C {sch, kel} of statistical systems based on the Kubo-Martin-
Schwinger (KMS) \C {mar, kubo} boundary condition for a field:
$$
\Phi (t)=\Phi (t-i\beta)
$$
This formal trick introduces into formalism the temperature $T=1/\b$ but,
without fail, leads to the $equilibrium$ fluctuation-dissipation conditions
\C {haag} (see also \C {chu}). Beside this we should have the two-temperature
theory describing kinetic energy dissipation process (for initial and final
states separately). It is evident that in such theory with two temperatures
it is impossible to use the KMS boundary condition.

In the $S$-matrix approach finite-temperature description can be introduced
(e.g. \C {kaj} and references cited therein) taking into account
that, for instance,
$$
d \Gamma_n =|a_{n,m}| \prod_{1}^{n}\frac{d^3 q_i}{(2\pi )^3 2\epsilon
(q_i)} , \;\;\; \epsilon (q)= (q^2 +m_h^2)^{1/2},
$$
is  the differential measure of final state. Then we can define the
temperature as the function of initial energy through the equation of state,
i.e. proportional to the mean energy of created particles. Such introduction
of temperatures as the Lagrange multiplier is obvious for microcanonical
approach \C {mar}. The initial-state temperature will be introduced by the
same way. Using standard terminology \C{5}, we  will deal with the
`mechanical' perturbations only \C{6} and it will not be necessary to
divide the perturbations on `thermal' and `mechanical' ones \C{int}.

Introducing temperature as the Lagrange multiplier we should assume that the
temperature fluctuations are small (Gaussian). In opposite case the notion
of temperature looses its sense. The `working' idea concerning nonequilibrium
processes based on assumption that evolution of a system goes through few
phases. In the first `fast' phase the $s$-particle distribution functions ${\bf D}_s$, $s>1$, strongly depends from initial conditions. But at the end
of this phase the system forgets the initial-state information. Second phase
is the `kinetic' one. One can expect that the space-time fluctuations of
thermodynamical parameters in this phase are large scale, i.e. there is
macroscopical domains in which the subsystems are equilibrium, with Gaussian
fluctuations of thermodynamical parameters. In the last `hydrodynamical'
phase the whole system is described by macroscopical parameters. We will see
that the Schwinger-Keldysh \C {sch, kel, land} formalism is applicable for
`hydrodynamical' phase only.

The above described $S$-matrix finite-temperature description can be realized
not only for uniform temperature distribution (we have done first step in
this direction wishing to introduce initial and final temperatures
separately). So,
introducing cells of measuring device (calorimeter) and introducing the
energy-momentum shells of each cell separately we can introduce the
individual temperatures in each cell. This can be done since in the
$S$-matrix theory the measurement performed by free (mass-shell) particles,
i.e. the measurement of energy (and momentum) can be performed in each cell
separately. This allows to capture the `kinetic' phase also (if the number
of calorimeter cells is high enough). In this phase multiparticle
distribution functions ${\bf D}_s$, $s>1$, are functionals of one-particle
distribution function ${\bf D}_1$ only. This means the `shortened'
description of the nonequilibrium medium \C{bog}. We will return to this
question in Sec.4 considering the dissipative processes thermal descriptions
applicability range.

The microcanonical description assumes that the energy of system is known
with arbitrary accuracy. Introducing the measurement cells and corresponding
energy shells we assume that the energy in each cell can be measured with
arbitrary accuracy. It is why we should work in the frame of Wigner functions
formalism \C{carr}.

E.~Wigner had offered the function $W(q,R)$ for the quantum states phase
space description \C{wig}:
$$
W(q,R)=\int dr e^{iqr}
\Psi (R+r/2) \Psi^* (R-r/2),
$$
where $\Psi (x)$ is the wave function of state. The existence of other
approaches must be mentioned \C{*}. But as will be seen below the Wigner's
description is mostly natural for us.

In the classical limit $\hbar =0$ the function $W(q,r)$ coincides with the
phase space probability distribution function. It obeys the equation \C{liu}:
$$
\dot{W}=\{W,H\}+O(\hbar),
$$
which coincides with Liouville equation only in the classical limit
$\hbar =0$.

The extension of Wigner's idea  on the relativistic case uses the connection
between Wigner's approach and inclusive description of inelastic scattering
processes \C{carr, hu}. But the Wigner functions are not directly measurable
quantities because of the quantum uncertainty principle $\D q \D r \sim
\hbar$. Just this restriction leads to impossibility to take the measurement
(calorimeter) cells 4-dimension $\Delta r$ arbitrary small and defines the
natural boundary of Wigner functions approach applicability. Wishing to use
the Wigner-functions description of experiments the corresponding theory must
take into account this restriction. The discussion of this question is
given in Sec.4.

So, in our terms one can use the thermodynamical formalism if the
nonstationary mediums `shortened' description may be applied: in this case
a mean value of correlation functions over the space-time are negligible
and the fluctuations of thermodynamical parameters are small (Gaussian).
Other approach should be mentioned also. Proposing \C{zub} that the
equilibrium in the nonuniform nonstationary medium may be attained in
small regions more quickly than in the whole system the entropy
maximalness in this restricted domains of a system can be used for
construction of the `local equilibrium density matrix' (LDM) \C{zub}. But
LDM is applicable for description of processes in which dissipation may be
disregarded \C{paz}. Nevertheless, if the energy-momentum density of
nonstationary
flow is considerably smaller than the energy density of matter then first
one can be taken into account perturbatively considering LDM as the
initial condition. This modifies LDM to `nonstationary density matrix'
(NDM) of Zubarev \C{zub} introducing an infinitesimal interaction with a
heat bath to get the increasing entropy. We will return to this question
in Sec.5.

The $S$-matrix will be introduced phenomenologically, using ordinary in a
quantum field theory reduction formalism. This leads naturally to necessity
to introduce the boundary conditions for interacting fields $\P (\s_{\infty})$,
where $\s_{\infty}$ is the infinitely far hypersurface, e.g. \C{bog1}.
The value of $\P (\s_{\infty})$ specify the environment of a system.

We start from vacuum boundary condition $\P(\s_{\infty})=0$ familiar for
a field  theory. This theory can be applied in the particle physics. The
simplest choice of $\P (\s_{\infty})\neq 0$ assumes that the system under
consideration is surrounded by black-body radiation. Just this `boundary
condition' restores the Schwinger-Keldysh \C{land} real-time
finite-temperature field theory \C {sem} from $S$-matrix  formalism in the
`hydrodynamical' phase and gives the dynamical interpretation of
the KMS periodic boundary condition.

One should admit also that last choice of boundary condition is not
unique: one can consider another organization of the environment of
considered system. The $S$-matrix interpretation is able to show the way
of adoption of formalism to the arbitrary environment\footnote{This
question was considered also in \C{hu}}. It should broaden the
potentialities of the real-time finite-temperature field-theoretical
methods. For instance, for heavy nucleus high energy interactions. The
special interest represent also the topological effects, but, by above
mention reason, in this paper consideration will be performed in the
perturbation theory framework only (see also concluding Section).

The central purpose of this review paper is to describe connections between
ordinary $S$-matrix description and popular in the modern literature
real-time finite-temperature field theories. We wish to discuss:\\
	--The QCD jets dominance in deep asymptotics over $n$ (Sec.2);\\
In this section we would like to show why the $real-time$ formalism is needed
for our dissipative process description.\\
	--The $S$-matrix interpretation of Schwinger-Keldysh theory (Sec.3);\\
In this section the uniform temperature description of the state will be
introduced in the spirit of microcanonical description. It is shown the way
of explicit calculations to show coincidence of used microcanonical
description and ordinary (Gibbs) canonical formalism. \\
	--The range of applicability of finite-temperature description
(Sec.4);\\
In this section the necessity and sufficiency of Bogolyubov's `shortened'
description is discussed. \\
	--The $S$-matrix description of media with nonuniform temperature
distribution (Sec.5);\\
In this section the Wigner functions formalism is introduced. The range if
its applicability to describe an experiment is shown.\\
	--The comparison of our $S$-matrix approach with `nonstationary
statistical operator' of Zubarev \C{zub} (Sec.6).\\
In this section the main distinction between $S$-matrix (microcanonical) and
Zubarev's (canonical) perturbation theories is shown.\\
	--Concluding remarks (Sec.7).\\
In this section the way as the nonperturbative effects may be included in the
formalism is discussed.

\section{Phenomenology}\0

To build the phenomenology \C{sys} of high multiplicity processes let us
introduce the classification of asymptotics over $n$. For this purpose it
is useful to consider the `big partition function':
$$
T(z,s)=\sum_n z^n \s_n (s), ~~~T(1,s)=\s_{tot} (s).
$$
Strictly speaking, summation over $n$ is performed up to $n_{max}$. But we
can extend summation up to infinity\footnote{I.e. wishing to consider the
`thermodynamical limit'.} if the weight $z$ is sufficiently small,
$0<z<z_{max}$. So, $T(z,s)$ can be considered as the nontrivial function of
$z$ with sufficient accuracy. Note that $z_{max}>1$ since $\s_n (s)$ decrease
with $n$.

If we know $T(z,s)$ then $\s_n (s)$ is defined by inverse Mellin
transformation. This gives (usual in thermodynamics) equation (of state):
\be
n=z\f{\pa}{\pa z}\ln T(z,s)
\l{1}\ee
Solving this equation we can estimate the asymptotics of $\s_n$:
\be
\s_n (s) \sim e^{-n \ln \bar{z} (n,s)},
\l{2}\ee
where $1<\bar{z} (n,s)<<z_{max}$ is smallest solution of eq.(\r{1}).

It follows from (\r{2}) that at $n\rightarrow \infty$ the solution of (\r{1})
must tend to singularity $z_s$ of $T(z,s)$ and the character of singularity
is not important. So, we must consider three possibility:
$$
a).~z_s=z_a=1,~~~b).~z_s=z_b=\infty,~~~c).~z_s=z_c,~~1<z_c < \infty.
$$
Following to Lee and Yang \C{lee} there is not singularities at $0<z<1$.

Let us consider now the physical content of this classification.
\\   a) $z_s=1$. \\
It is known that the singularity $z_s=1$ reflects the first
order phase transition \C{lee}. To find $\s_n$ for this case we would
adopt Langer's analyses \C{lan}. Introducing the temperature $1/\b$ instead
of total energy $\sqrt{s}$ we can use the isomorphism with Ising model.
For this purpose we divide the space volume on cells and if there
is particle in the cell we will write (-1). In opposite case (+1). It is
the model of lattice gas well described by Ising model. We can regulate the
number of down-looking spins, i.e. the number of created particles, by the
external magnetic field ${\bf H}$. Therefore, $z=\exp\{-\b{\bf H}\}$ and
${\bf H}$ is the chemical potential.

The corresponding partition function in the continuous limit \C{lan} (see
also \C{kac2}) has the form:
\be
R(\b,z)=\int D\mu e^{-\int dx \{\f{1}{2}(\vec{\pa}\mu )^2 -\e \mu^2 +
\a \mu^4 - \lambda \mu \}},
\l{3} \ee
where $\e \sim (1-\f{\b_c}{\b})$ and $\lambda \sim {\bf H}$,
with critical temperature $1/\b_c$.

If $\b_c >\b$ there is not phase transition and the potential has one
minimum at $\mu =0$. But if $\b_c <\b$ there is two degenerate minimum
at $\mu_{\pm}=\pm \sqrt{\e/2\a}$ if $\lambda =0$. Switching on ${\bf H}<0$
the left minimum at $\mu_{-}\sim -\sqrt{\e/2\a}$ becomes absolute and the
system will tunnel into this minimum (see also \C{col}). This process
describes particles creations as a process of spins flippings.

The eq.(\r{1}) gives at $n\rightarrow \infty$
$$
\ln \bar{z}\sim n^{-1/3} > 0.
$$
In result,
$$
\s_n \sim e^{-an^{2/3}} >0(e^{-n}),~~~a>0,
$$
i.e. decrease slower then $e^{-n}$. The quasiclassical calculation shows
that the functional determinant is singular at ${\bf H}=0$. It must be
underlined that in the used Ising model description the chemical
potential deforms ground state. In result the quasiclassical approximation
is applicable since $\ln \bar{z}<<1$, i.e. since the processes of spin
flippings are rear at high multiplicity region. It is easy to show in
this approximation \C{lan} that the functional determinant is singular at
${\bf H}=0$, i.e. at $z=1$.

Note that $\bar{z}$ $decrease$ to one with $n$. This unusual phenomena must
be explained. Considered above mechanism of particles creation describes
`fate of false vacuum' \C{col}. In the process of decay of unstable
state the clusters of new phase of size $X$ are created. If the
cluster have dimension $X>X_c$ its size increase since the
volume energy ($\sim X^3$) of the cluster becomes better then the surface
tension energy ($\sim X^2$). This condition defines the value of $X_c$.
The `critical' clusters wall accelerate, i.e the work needed to add one
particle into cluster decrease with $X>X_c$. This explains the reason why
$\bar{z}$ decrease with $n$ noting that $\ln \bar{z}$ is proportional to
Gibbs free energy per one particle.

The described mechanism of particles creation assumes that we had prepared
the $equilibrium$ system in the unstable phase at $\mu_{+}\sim +\sqrt{\e/2\a}$
and going to another state at $\mu_{-}\sim -\sqrt{\e/2\a}$ the system creates
particles. The initial state may be the QGP and final state may be the hadrons
system. Therefore, we must describe the way as the quarks system was prepared.

Following to Lee-Yang's picture of first order phase transition \C{lee}(see
also \C{kac2}) there is not phase transition in a finite system
(the partition function can not be singular for finite $n_{max}$). This
means that the multiplicity (and the energy) must be high enough to see
described phenomena.
\\   b) $z_s =\infty$.\\
Let us return to the integral (\r{3}) to investigate the case $\b_c>\b$. In
this case the potential has one minimum at $\mu =0$. The external field
${\bf H}$ creates the mean field $\bar{\mu}=\bar{\mu}({\bf H})$ and the
integral (\r{3}) should be calculated expanding it near $\mu=\bar{\mu}$.
In result, in the quasiclassical approximation ($\bar{\mu}$ increase with
increasing $n$),
$$
\ln R(\b ,z) \sim (\ln z)^{4/3}.
$$
This gives $\ln \bar{z} \sim n^3$ and $\ln \s_n \sim -n^4$, i.e.
$$
\s_n < 0(e^{-n}).
$$

There is also other possibility to interpret considered case b). For this
case we can put
\be
\ln T(z,s)=n_0 (s) + \bar{n}(s)(z-1) + O((z-1)^2)
\l{4}\ee
at $|z-1|<<1$. By definition $n_0 (s)=\ln \s_{tot}$.
The experimental distribution of $\ln T(z,s)-n_0 (s)$ for various
energies shows that the contributions of $O((z-1)^2)$ terms increase
with energy \C{epan}. The hadrons `standard model' (SM) assumes that
$$
\ln t(z,s)=n_0 (s) + \bar{n}(s)(z-1)
$$
is the Born term in the perturbation series (\r{4}). There is various
interpretations of this series, e.g. the multiperipheral model, the Regge
pole model, the heavy color strings model, the QCD multiperipheral models,
etc. In all this models $n_0 =a_1 +a_2 \ln s$, $0\leq a_2 <<1$ and
$\bar{n}(s)=b_1 +b_2 \ln s$, $b_2 >0$. The second ingredient of hadrons
SM is the assumption that mean value of created particles transfers momentum
$<k>=const$, i.e. is the energy (and multiplicity) independent. It can be
shown that under this assumptions:
\be
\ln T(z,s)=n_0 (s) + \sum_{n} c_n (s) (z-1)^n,~~~c_1 \equiv \bar{n}
\l{5}\ee
is $regular$ at finite values of $z$ \C{epan} and is able to give well
confirmed by experiment predictions.

Inserting (\r{5}) into (\r{1}) we find that $\bar{z} (n,s)$ is the increasing
function of $n$. Therefore,
\be
\s_n < O(e^{-n}).
\l{7}\ee

But the SM have a finite range of validity: beyond $n\sim \bar{n}^2$ the
model must be changed since it is impossible to conserve $<k>=const.$ at
higher multiplicities \C{grib}.

We should underline once more that only two possibilities a) and b) can be
deduced from representation (\r{3}), see also \C{lan}. But nevertheless
there is other possibility:
\\   c) $1<z_s<\infty$.\\
Let us assume now that
\be
T(z,s) \sim (1-\f{z-1}{z_c -1})^{-\ga},~~~\ga >0.
\l{6}\ee
Then, using normalization condition, $(\pa T(z,s)/\pa z)|_{z=1}=\bar{n}_j
(s)$ we can find that $z_c (s)=1+ \ga/\bar{n}_j (s)$. The singular
structure (\r{6}) is impossible in SM because of condition $<k>=const$.
But if $|z-1|<<1$ we have estimation (\r{4}). The difference between SM
and c) is seen only at $1-(z-1)/(z_c -1)<<1$, i.e. or in asymptotics over
$n$ or in asymptotics over energy. The singular structure is familiar for
`logistic' equations of QCD jets, e.g. \C{jet}.

In considered case  $\bar{z}=z_c +0(\bar{n}_j /n)$ and at high energies
($\bar{n}_j (s)>>1$)
\be
\s_n \sim e^{- \ga n/\bar{n}_j} = O(e^{-n}).
\l{12}\ee

Therefore,

{\it comparing (\r{7}) and (\r{12}) we can conclude that at sufficiently high
energies, i.e. if $\bar{n}_j>>\bar{n}$, where $\bar{n}$ is the SM mean
multiplicity, the mechanism c) must dominate in asymptotics over $n$}.

It is the general, practically model independent, prediction. It has
important from experimental point of view consequence that at high energies
there is wide range of multiplicities where the SM mechanism of hadrons
creation is negligible. In other words, the CQGP of high multiplicity
processes is the dynamical consequence of jets and SM mechanisms. At
transition region between `soft' of SM and `hard' of jets one can expect
the `semihard' processes of minijets dominance.

The multiplicity distribution in jets has interesting property noted many
decades ago by Volterra in his mathematical theory of populations \C{vol}.
In our terms, if one-jet
partition function has the singularity at $z_c^{(1)} (s)=1 +
\ga /\bar{n}_j (s)$ then two-jet partition function must be singular at
$$
z_c^{(2)} (s)=1+\f{\ga}{\bar{n}_j (s/4)} > z_c^{(1)} (s),
$$
and so on. Therefore, at high energies and $n>\bar{n}_j (s)$ the jets number
must be minimal (with exponential accuracy). This means that at $n\rightarrow
\infty$ the processes of hadrons creation have a tendency to be Markovian
(with sharp increase of transverse momentum $<k>$) and only in the last stage
the (first order) phase transition (colored plasma) $\rightarrow$ (hadrons)
may be seen.

One can say that in asymptotics over $n$ we consider the `inflational'
chanel of thermalization which is so fast\footnote{The partons life time
with virtuality $|q|$ is $\sim 1/|q|$ and the time needed for hadrons
of mass $m_h$ formation is $\sim 1/m_h$. Therefore the parton have a time to
decay before hadrons formation if $|q| >> m$. But this situation is rear
since the thermal motion in the initial stage of process is high.} that the
usual confinement forces are `freezed' and do not play important role
in final colored plasma creation.

\section {$S$-matrix interpretation} \0

	{\it 3.1. Vacuum boundary conditions.}

The starting point of our calculations is $n$- into $m$-particles
transition amplitude $a_{n,m}$, the derivation of which is well
known procedure  in the perturbation theory framework. For this purpose the
$(n+m)$-point Green function $G_{n,m}$ is introduced \C{vas}. To calculate
the nontrivial elements of $S$-matrix one
must put the external particles on the mass shell. Formally this
procedure means amputation of the external legs of $G^{c}_{n,m}$
and further multiplication on the free particles wave functions.
In result the amplitude of $m$- into $n$-particles transition $a_{n,m}$
in the momentum representation has the form:
\be
a_{n,m}((q)_n;(p)_m)=(-i)^{n+m}\prod_{k=1}^{m}\hat{\phi}(q_k)
\prod_{k=1}^{n}\hat{\phi}^* (p_k) Z(\phi).
\l{9}
\ee
Here we introduce the `annihilation' operator
\be
\hat{\phi}(q)=\int dx e^{-iqx} \hat{\phi}(x),\;\;\;
\hat{\phi}(x)=\frac{\delta}{\delta \phi (x)},
\l{13}
\ee
$\hat{\phi}^* (p_k)$ is the `creation' operator and $q_{k}$ and $p_{k}$ are
the momentum of in- and out-going particles. In (\r{9})
$$
Z(\phi)=\int D\Phi e^{iS(\Phi)-iV(\Phi+\phi)}
$$
is the generating functional. The total action was divided on
two parts, where $S(\P)$ is the free part and $V(\P ,\p)$ describes the
interactions. At the very end one should put the auxiliary field $\p =0$.

To provide the convergence of the integral (\ref {9}) over scalar field
$\P$ the action $S(\P)$ must contain positive imaginary part. Usually for
this purpose Feynman's $i\e$-prescription is used. It is better for us to
shift infinitesimally time contour to the upper half plane \C{mil,land}, i.e.
to the Mills contour
$$
C_+ :t\rightarrow t+i\e, \;\;\; \e >0
$$
and after all calculations to return the time contour on the real axis,
$\e \rightarrow +0$.

In eq. (\ref{9}) the integration is performed over all field
configurations with standard vacuum boundary condition:
$$
\int d^4 x \partial_{\mu}(\Phi \partial^{\mu}\Phi)=
\int_{\sigma_{\infty}}d\sigma_{\mu}\Phi\partial^{\mu}\Phi=0,
$$
which assumes zero contribution from the surface term.

Supposing that the particles number and momenta are insufficient for us
we introduce the probability
\be
r(P)=\sum_{n,m}\frac{1}{n!m!}\int d\omega_n (q) d\omega_m (p)
\delta^{(4)}(P-\sum_{k=1}^{n} q_k)
\delta^{(4)}(P-\sum_{k=1}^{n} p_k) |a_{n,m}|^2.
\l{10}
\ee
where
$$
d\omega_n(q)=\prod_{k=1}^{n}d\omega(q_k)=
\prod_{k=1}^{n}\frac{d^3  q_k}{(2\pi)^3 2\epsilon (q_k)}, \;\;\;\;
\epsilon =(q^2+m_h^2)^{1/2},
$$
is  the Lorentz-invariant phase space element. We assume that the
energy-momentum conservation $\d$-function was extracted from the
amplitude. It was divided onto two parts:
\be
\delta^{(4)}(\sum q_k - \sum p_k)=\int d^4 P
\delta^{(4)}(P-\sum q_k)\delta^{(4)}(P-\sum p_k)
\l{de}\ee
It is not too hard to see that, up to phase space volume,
$$
r=\int d^4P r(P)
$$
is the imaginary part of amplitude $<vac|vac>$. Therefore, computing
$r(P)$ the  standard renormalization procedure  can be  applied and the
new divergences will not arise in our formalism.

The Fourier  transformation of $\d$-functions in (\ref{10}) allows to
write $r(P)$ in the form:
$$
r(P)=\int \frac{d^4 \alpha_1}{(2\pi)^4}\frac{d^4\alpha_2}{(2\pi)^4}
e^{iP(\alpha_1+\alpha_2)}
\R(\alpha_1, \alpha_2),
$$
where
\be
\R(\alpha_1, \alpha_2)=
\sum_{n,m} \frac{1}{n!m!}\int
\prod_{k=1}^{n}\{d\omega(q_k)e^{-i\alpha_1 q_k}\}
\prod_{k=1}^{m}\{d\omega(p_k)e^{-i\alpha_2 p_k}\} |a_{n,m}|^2.
\l{11}\ee
Introduction of the `Fourier-transformed' probability $\R(\a_1 ,\a_2 )$
means only that the  phase-space volume is not fixed exactly, i.e.
it is proposed that 4-vector $P$ is fixed with some accuracy if $\a_i$
are fixed.
The energy and momentum in our approach are still locally conserved
quantities since the amplitude $a_{nm}$ is translational invariant.
So, we  can perform the transformation:
$$
\a_1 \sum q_k =(\a_1 -\sigma_1 )\sum q_k +\sigma_1 \sum q_k
\rightarrow (\a_1 -\sigma_1 )\sum q_k +\sigma_1 P
$$
since 4-momenta are conserved. The choice of $\sigma_1$ fixes the
reference  frame. This  degree of freedom of the theory was considered
in \C{mta, psf}.

Inserting (\ref{9}) into (\ref{11}) we find  that
\ba
\R(\alpha_1, \alpha_2)=\exp \{i\int dx dx'(
\hat{\phi}_+(x)D_{+-}(x-x',\alpha_2)\hat{\phi}_-(x')-
\n \\
-\hat{\phi}_-(x)D_{-+}(x-x',\alpha_1)\hat{\phi}_+(x'))\}
Z(\phi_+)Z^* (\phi_-),
\l{8}\ea
where $D_{+-}$ and $D_{-+}$ are  the positive and negative frequency
correlation functions:
$$
D_{+-}(x-x',\alpha)=-i\int d\omega(q)e^{iq(x-x'-\alpha)}
$$
describes the process of particles creation at the  time  moment $x_0$
and its absorption at $x'_0$, $x_0>x'_0$, and $\a$ is the
center of mass () 4-coordinate. Function
$$
D_{-+}(x-x',\alpha)=i\int d\omega(q)e^{-iq(x-x'+\alpha)}
$$
describes the opposite process, $x_0<x'_0$. These functions obey the
homogeneous equations:
$$
(\partial^2 +m^2)_x G_{+-}=
(\partial^2 +m^2)_x G_{-+}=0
$$
since the propagation of mass-shell particles is described.

We suppose that $Z(\p)$ may be computed perturbatively. For this
purpose following transformation will be used:
\ba
e^{-iV(\phi)}=
e^{-i\int dx \hat{j}(x)\hat{\phi}'(x)}
e^{i\int dx j(x)\phi (x)}
e^{-iV(\phi ')}=
\n\\
=e^{\int dx \phi(x)\hat{\phi}'(x)}
e^{-iV(\phi ')}=
\n\\
=e^{-iV(-i\hat{j})}
e^{i\int dx j(x)\phi (x)},
\l{14}\ea
where $\hat{\phi}$ was defined in (\ref{13}). At the end of calculations
the auxiliary variables $j$, $\p'$ should be taken equal to zero. Using the
first equality in (\ref{14}) we find that
\be
Z(\phi)=
e^{-i\int dx \hat{j}(x)\hat{\Phi}(x)}
e^{-iV(\Phi+\phi)}
e^{-\frac{i}{2}\int dx dx'
 j(x)D_{++}(x-x')j(x')},
\l{15}
\ee
where $D_{++}$ is the causal Green function:
$$
(\partial^2 +m^2)_x G_{++} (x-y)=\delta (x-y)
$$
Inserting (\ref{15}) into (\ref{8}) after simple manipulations with
differential operators, see (\ref{14}) we find the expression:
\ba
\R(\alpha_1, \alpha_2)=
e^{-iV(-i\hat{j}_+)+iV(-i\hat{j}_-)}\times
\n \\ \times
\exp\{ \frac{i}{2} \int dx dx'(
 j_+ (x)D_{+-}(x-x',\alpha_1)j_- (x')-
\n \\
 j_- (x)D_{-+}(x-x',\alpha_2)j_+ (x')-
\n\\
- j_+ (x)D_{++}(x-x')j_+ (x')+
 j_- (x)D_{--}(x-x')j_- (x'))\},
\l{19}
\ea
where
$$
D_{--}=(D_{++})^*
$$
is the anticausal Green function.

Considering the system with large number of particles we
can simplify calculations choosing the CM frame $P=(P_0 =E,\vec 0)$.
It is useful also \C {kaj, mar} to rotate the contours of integration
over $\alpha_{0,k}$: $\alpha_{0,k}=-i\b_k, Im\b_k =0, k=1,2$.
In result, omitting unnecessary constant, we will consider
$\R=\R(\b_1,\b_2)$.

External particles play the double role in the $S$-matrix approach:
their interactions create and annihilate the interacting fields system
and, on the other hand, they are probes through which the
measurement of the system is  performed. Since $\b_k$ are the conjugate
to the particles  energies quantities we will interpret them
as  the inverse temperatures in the initial ($\b_+$) and  final ($\b_-$)
states of interacting fields. But there is the question: are constants
$\b_k$ really the  `good' parameters to describe the system.

The integrals over $\b_k$:
\be
r(E)=\int \frac{d\beta_1}{2\pi i}\frac{d\beta_2}{2\pi i}
e^{(\beta_1 +\beta_2)E}
e^{-F(\beta_1,\beta_2)},
\l{18}
\ee
where
$$
F(\beta_1,\beta_2)=-\ln \R(\beta_1,\beta_2),
$$
can be computed by the stationary phase method. This assumes that the total
energy $E$ is a fixed quantity. The solutions of the equations (of state):
\be
E=\frac{\partial F(\beta_1,\beta_2)}{\partial \beta_k} ,\;\;\;   k=1,2,
\l{17}
\ee
gives the mostly probable values of $\b_k$ at a given $E$. Eqs. (\ref{17})
always  have the real solutions and, because of energy conservation law,
both eqs. (\ref{17}) have the same solution with the  property \C {mar}:
$$
\beta_k=\beta (E), \;\;\;\;    \beta>0.
$$
Assuming that
$\b$ is the `good' parameter, i.e. the  fluctuations  of
$\b_k$ are Gaussian we can interpret $F(\b_1,\b_2)$
as  the free energy and $1/\b_k$ as the temperatures. Such definition
of thermodynamical parameters is in a spirit of microcanonical
description. We will return to this question in Sec.4.

The structure of generating functional (\ref{19}) is the same as the
generating functional of Niemi-Semenoff \C {sem} have. The difference is
only in the definition of Green functions which follows from the choice
of boundary condition (\ref{7}). The Green functions  $D_{ij}, i,j=+,-$
were defined on the time contours $C_{\pm}$ in the complex time
plane ($C_-=C_+^*$). This definition of the  time  contours coincide with
Keldysh' time contour \C {kel}. The expression (\ref{19}) can be written
in the compact form if the matrix notations are  used. Note also a
doubling of the  degrees of freedom. This doubling is unavoidable
since Green functions $D_{ij}$ are  singular  on the light cone.

	{\it 3.2. Closed-path boundary conditions.}

The generating functional $\R(\a_1,\a_2)$  has important factorized
structure, see (\r{8}):
$$
\R(\a_1,\a_2)=e^{\hat{N} (\a_1,\a_2;\p)}\R_0 (\phi_{\pm}),
$$
where the operator
\ba
\hat{N} (\a_1,\a_2;\p)=
\int dx dx'(\hat{\phi}_+(x)D_{+-}(x-x',\alpha_2)\hat{\phi}_-(x')-\n\\
-\hat{\phi}_-(x)D_{-+}(x-x',\alpha_1)\hat{\phi}_+(x'))
\ea
acts on the generating functional
\ba
\R_0 (\phi_{\pm})=Z(\phi_+)Z^* (\phi_-)=
\n \\
=\int D\Phi_+ D\Phi_-
e^{iS(\Phi_+)-iS(\Phi_-)-iV(\Phi_+ +\phi_+) + iV(\Phi_- +\phi_-)},
\l{20}
\ea
of measurables. All `thermodynamical' information was contained in the
operator $\hat{N} (\a_1,\a_2;\p)$ and interactions are hidden in
$\R_0 (\phi_{\pm})$. One can say that action of the operator $\hat{N}$ maps
the system of interacting fields on the measurable states. Last ones are
`labeled' by $\a_1$ and $\a_2$. Just this property allows to say that we
are dealing with `mechanical' fluctuations only. To regulate the particles
number we can introduce into $\hat{N}$ dependence from `activities' $z_1$
and $z_2$ for initial and final states separately.

The independent fields $\p_+,\p_-$ and $\P_+,\P_-$ were defined on the
time contours $C_+,C_-$. By definition, path integral (\ref{20}) describes
the closed path motion in the space of fields $\P$. We want to use this fact
and introduce a more general boundary condition which also guaranties
cancelation of surface terms in the perturbation  framework. We
will introduce the equality:
\be
\int_{\sigma_{\infty}} d\sigma_{\mu} \Phi_+  \partial^{\mu}\Phi_+ =
\int_{\sigma_{\infty}} d\sigma_{\mu} \Phi_-  \partial^{\mu}\Phi_-.
\l{21}
\ee
The solution of eq.(\ref{21}) requires that the fields $\P_+$ and $\P_-$
(and theirs first derivatives $\partial_{\mu}\P_{\pm}$) coincide on the
boundary hypersurface $\s_{\infty}$:
$$
\Phi_{\pm}(\sigma_{\infty})=\Phi(\sigma_{\infty}),
$$
where,  by definition, $\Phi(\sigma_{\infty})$
is the arbitrary, `turning-point', field.

The existence of nontrivial field $\Phi(\sigma_{\infty})$, in absence of
surface terms, has influence only on the structure of Green functions
\ba
G_{++}=<T\Phi_+\Phi_+>,\;\;\;
G_{+-}=<\Phi_+\Phi_->,
\n\\
G_{-+}=<\Phi_-\Phi_+>, \;\;\;
G_{--}=<\tilde{T}\Phi_-\Phi_->,
\l{}\ea
where $\tilde{T}$ is the antitemporal time ordering operator. This Green
functions must obey the  equations:
\ba
(\partial^2 +m^2)_x G_{+-} (x-y)=
(\partial^2 +m^2)_x G_{-+} (x-y)=0,
\n\\
(\partial^2 +m^2)_x G_{++} (x-y)=
(\partial^2 +m^2)_x^* G_{--} (x-y)=\delta (x-y),
\l{}\ea
and the general solution of these equations:
\ba
G_{ii}=D_{ii}+g_{ii},
\n \\
G_{ij}=g_{ij},\;\;\;\; i\neq j
\l{22}
\ea
contain the undefined terms $g_{ij}$ which must obey the homogenous equations:
\be
(\partial^2 +m^2)_x g_{ij} (x-y)=0,\;\;\; i,j=+,-.
\l{23}
\ee
The general solution of these equations (they are distinguished by the
choice of the time contours $C_{\pm}$)
\be
g_{ij}(x-x')=\int d\omega (q) e^{iq(x-x')} n_{ij} (q)
\l{24}
\ee
are defined by the functions $n_{ij}$. Last ones are the
functionals of `turning-point' field $\Phi(\sigma_{\infty})$: if
$\Phi(\sigma_{\infty})=0$ we must have $n_{ij}=0$ and we will come back
to the theory of previous section.

Our aim is to define $n_{ij}$. We can suppose that
$$n_{ij}\sim <\Phi(\sigma_{\infty})\cdots\Phi(\sigma_{\infty})>.$$
The simplest supposition gives:
\be
n_{ij}\sim <\Phi_{i}\Phi_{j}>\sim <\Phi^2(\sigma_{\infty})>.
\l{25}
\ee
We will find the exact definition of
$n_{ij}$ starting from the $S$-matrix interpretation of the  theory.

We should suppose there are only free, mass-shell, particles that on the
infinitely far hypersurface $\s_{\infty}$. Formally this follows from
(\ref{22}) -(\ref{24}) and is natural in the $S$-matrix framework \C{pei}.
In other respects choice of the boundary condition is arbitrary.

Therefore, our aim is the description of evolution of the
system in a background field of mass-shell particles. We will assume that
there are not any special correlations among background particles and will
take into account only the energy-momentum conservation laws constraints.
Quantitatively this means that multiplicity distribution of background
particles is Poison-like, i.e. is  determined by the mean multiplicity only.
This is in spirit of definition of $n_{ij}$ in eqs.(\ref{24}), (\ref{25}).

Our derivation is the same as in \C {psf}. Here we restrict ourselves
mentioning only the main quantitative points.

In the vacuum case of Sec.3.1 the process of particles creation and theirs
further absorption was described. In presence of the background particles
this time-ordered picture is wiped out: appears possibility of particles
absorption before theirs creation.

The particles creation and absorption was described by the product of operator
exponent (\r{8}). One can derive (see also \C {psf}) the generalizations of
(\ref{8}): presence of the background particles will lead to the same
structure:
$$
\R_{cp}=e^{i\hat{N}(\phi_i^*\phi_j)}\R_0(\phi_{\pm}),
$$
where $R_0 (\p_{\pm})$ is the same generating functional, see (\ref{20}).
But the operator $\hat{N}(\p^*_{i}\p_{j}), i,j=+,-,$ should be changed
wanting to take into account the external particles environment.

The operator $\hat\p^*_i(q)$ was interpreted as the creation and
$\hat\p_i(q)$ as the annihilation operator, see definition (\ref{9}).
Correspondingly the product $\hat\p^*_i(q)\hat\p_j(q)$ acts as
the activity operator. So, in the expansion of $\hat{N}(\p^*_i\p_j)$
we can leave only first nontrivial term:
\be
\hat{N}(\phi^*_i\phi_j)=
\int d\omega (q) \hat{\phi}^*_i (q) n_{ij} \hat{\phi}_j (q),
\l{26}
\ee
since  no special correlation among background particles should be expected.
If the external (nondynamical) correlations are present then the higher
powers of $\hat\p^*_i\hat\p_j$ will appear in expansion (\ref{26}) \C{hu}.
Following to the interpretation of $\hat\p^*_i\hat\p_j$ we conclude that
$n_{ij}$ is the mean multiplicity of background  particles. In (\ref{26})
the  normalization condition $N(0)=0$ was used and summation over all
$i,j$ was assumed. (In the vacuum case only the combinations $i\neq j$ was
present.)

Computing $R_{cp}$ we must conserve the translational invariance of
amplitudes and extract the energy-momentum conservation $\d$-functions.
We must adjust to each vertex of in-going particle in $a_{n,m}$ the factor
$e^{-i\a_1q/2}$ and for each out-going particle $e^{-i\a_2q/2}$ one after
Fourier transformation of this $\d$-functions.

So, the product $e^{-i\a_kq/2}e^{-i\a_jq/2}$ can
be interpreted as the probability factor of the one-particle
$(creation+annihilation)$ process. The $n$-particles
$(creation+annihilation)$ process' probability is the simple
product of  these factors if  there is not special correlations
among background  particles. This interpretation is evident in the
CM frame $\a_k=(-i\b_k,\vec0)$.

After this preliminaries it is not hard to find that in the
CM frame we have:
\be
n_{++}(q_0)=n_{--}(q_0)=
=\frac{1}{e^{\frac{\beta_1 +\beta_2}{2}|q_0|}-1} \equiv
\tilde {n}(|q_0|\frac{\beta_1 +\beta_2}{2}).
\l{27}
\ee
Computing $n_{ij}$ for $i\neq j$ we must take into account that we have one
additional particle:
\be
n_{+-}(q_0)=
= \Theta (q_0)(1+\tilde {n}(q_0 \beta_1))+
 \Theta (-q_0)\tilde {n}(-q_0 \beta_1)
\l{**}\ee
and
\be
n_{-+}(q_0)=
 \Theta (q_0)\tilde {n}(q_0 \beta_2)+
 \Theta (-q_0)(1+ \tilde {n}(-q_0 \beta_2)).
\l{28}
\ee
Using (\ref{27}), (\ref{**}) and (\ref{28}), and
the definition (\ref{22}) we find the Green functions (the
matrix Green functions in the real-time finite-temperature field theories
was introduced firstly in \C{magr}):
$$
G_{i,j}(x-x',(\beta))=\int \frac{d^4 q}{(2\pi)^4} e^{iq(x-x')}
\tilde{G}_{ij} (q, (\beta))
$$
where
\ba
i\tilde{G}_ij (q, (\beta))=
\left( \matrix{
\frac{i}{q^2 -m^2 +i\epsilon} & 0 \cr
0 & -\frac{i}{q^2 -m^2 -i\epsilon} \cr
}\right)
+\n \\ \n \\+
2\pi \delta (q^2 -m^2 )
\left( \matrix{
\tilde{n}(\frac{\beta_1 +\beta_2}{2}|q_0 |) &
\tilde{n}(\beta_2 |q_0 |)a_+ (\beta_2) \cr
\tilde{n}(\beta_1 |q_0 |)a_- (\beta_1) &
\tilde{n}(\frac{\beta_1 +\beta_2}{2}|q_0 |) \cr
}\right)
\l{31}
\ea
and
$$
a_{\pm}(\beta)=-e^{\frac{\beta}{2}(|q_0|\pm q_0)}.
$$
The corresponding generating functional has the standard form:
\ba
\R_{cp}(j_{\pm})=\exp\{-iV(-i\hat{j}_+)+iV(-i\hat{j}_-)\}\times
\n \\ \times
\exp\{\frac{i}{2}\int dx dx' j_i (x)G_{ij}(x-x',(\beta))j_j(x')\}
\l{29}
\ea
where the summation over repeated indexes  is assumed.

Inserting (\ref {29}) in the equation of state (\ref{17}) we can find that
$\beta_1 =\beta_2 =\beta (E)$. If $\beta (E)$ is a  `good' parameter then
$G_{ij}(x-x';\beta )$ coincide with the Green functions of
the real-time finite-temperature field theory and the KMS boundary
condition:
\be
G_{+-}(t-t')=G_{-+}(t-t'-i\beta),\;\;\;
G_{-+}(t-t')=G_{+-}(t-t'+i\beta),
\l{30}
\ee
is restored. The eq.(\ref{30}) can be deduced from (\ref{31}) by
the direct calculations.

\section{Applicability of finite-temperature description}\0

	{\it 4.1. The Schwinger-Keldysh formalism.}

There is various approaches to build the real-time finite-temperature field
theories of Schwinger-Keldysh type (e.g. \C{land}). All of them uses
various tricks for analitical continuation of imaginary-time Matsubara
formalism to the real time \C{kad}. The basis of the approaches is
introduction of Matsubara field operator
\be
\P_M ({\bf x}, \b)=e^{\b H}\P_S ({\bf x} e^{-\b H},
\l{32}\ee
where $\P_S ({\bf x})$ is the interaction-picture operator, instead of
Heisenberg operator
$$
\P ({\bf x}, t)=e^{it H}\P_S ({\bf x} e^{-it H}.
$$
This introduces the averaging over Gibbs ensemble instead of averaging
over zero-temperature vacuum states.

If the interaction switched on adiabatically at the instant $t_i$ and
switched off at $t_f$ then there is the unitary transformation:
$$
\P (x)=U(t_i ,t_f )U(t_i ,t)\P_S (x)U(t ,t_i).
$$
Introducing the complex Mills time contours \C{mil} to connect $t_i$ to
$t$, $t$ to $t_f$ and $t_f$ to $t_i$ we form `closed-time' contour $C$
(the end-points of the contours $C_+$ and $C_-$ are
$joint$ together). This allows to write last equality in the compact form:
$$
\P (x) =T_C \{ \P (x) e^{i\int_C d^4 x' L_{int} (x')} \}_S,
$$
where $T_C$ is the time-ordering on the contour $C$ operator.

The corresponding expression for the generating functional $Z(j)$ of
correlation (Green) functions has the form:
$$
Z(j)=R(0)< T_C  e^{i\int_C d^4 x \{L_{int} (x) + j(x)\P (x)\}_S}>,
$$
where $<>$ means averaging over initial state.

If the initial correlations have little effect we can perform averaging over
Gibbs ensemble. This is the main assumption of formalism: the generating
functional of the Green functions $Z(j)$ has the form in this case:
$$
Z(j)=\int D\P' <\P';t_i| e^{-\b H} T_C e^{i\int_C d^4 x  j(x)\P (x)}|\P'; t_i>
$$
with $\P' =\P' ({\bf x})$. In accordance with (\r{32}) we have:
$$
<\P';t_i| e^{-\b H} = <\P';t_i -i\b|
$$
and, in result,
\be
Z(j)=\int D\P' e^{i\int_{C_{\b}} d^4 x \{ L(x) + j(x)\P (x)}
\l{33}\ee
where path integration performed with KMS periodic boundary condition:
$$
\P (t_i) = \P (t_i -i\b).
$$
In (\r{33}) the contour $C_{\b}$ connects $t_i$ to $t_f$, $t_f$ to $t_i$
and $t_i$ to $t_i -i\b$. Therefore it contains imaginary-time Matsubara
part $t_i$ to $t_i -i\b$. More symmetrical formulation uses following
realization: $t_i$ to $t_f$, $t_f$ to $t_f - i\b/2$, $t_f - i\b/2$ to
$t_i - i\b/2$ and $t_i -i\b/2$ to $t_i -i\b$ (e.g. \C{sem}). This case
also contains the imaginary-time parts of time contour. Therefore,
eq. (\r{33}) presents the analitical continuation of Matsubara generating
functional to real times.

One can note that if this analitical continuation is possible in $Z(j)$
then representation (\r{33}) gives good recipe of regularization of
frequency integrals in the Matsubara perturbation theory, e.g. \C{land},
but nothing new for
our problem since the Matsubara formalism is a formalism for equilibrium
states only.

Taking $t_i =-\infty$ and $t_f =+\infty$ and calculating integral (\r{33})
perturbatively we find coincidence of $Z(j)$ and $R(\b)$ from (\r{29}) with
Green functions defined in (\r{31}) $if$ $\b_1 =\b_2$. This `factorization'
of contributions from contours $C_+$ and $C_-$
in the integral (\r{33}) follows from Rieman-Lebesque lemma \C{leb} which
is applicable in the perturbation framework \C{mil, sem}. Note absence of
Matsubara parts of contour, which prevents the factorization, in the derived
`$S$-matrix generating functional' (\r{29}) by definition (importance of this
circumstances is discussed in Sec.7).

	{\it 4.2. Range of the `hydrodynamical' approximation.}

Let us return now to eq.(\r{11}). To find the physical meaning $\b_{1(2)}$
we must show the way as they can be measured. If there is nonequilibrium
flow it is hard to invent a thermometer (or thermodynamical calorimeter)
which measures locally in space-time the temperatures of this dissipative
processes. But there was described another way - to define the
temperatures through equations of state. This is possible in the
accelerator experiments where the total energy $E$ is fixed. So, we will
define $\b_{1(2)}$ through equations of state (\r{17}), i.e. considering
$1/ \b_{1(2)}$ as the mean energy of particles in the initial
(final) state. But even knowing solutions of this equations one can not find
$\R(E,z)$ correctly if the assumption $\b_{1(2)}$ are `good' quantities
is not added, i.e. that the fluctuations near solutions of eqs.(\r{17}) are
small (Gaussian).

This assumption is the main problem toward nonequilibrium thermodynamics. The
problem in our terms looks as follows: the expansion near $\b_{1(2)}(E)$
gives asymptotic series over
$$ \int {\bf D}_s \sim
\int \prod \{d\o (k_i) dr_i\} <\e (k_1) \e (k_2)\cdots >|_{(r_1,r_2,...)},
$$
where $<>_{()}$ means averaging over fields drown on fixed points of phase
space $(k,r)_i$. In other words, the fluctuations near $\b_{1(2)}(E)$ are
defined by value of inclusive spectra familiar in particles physics.
Therefore, $\b_{1(2)}(E)$ are `good' quantities if this inclusive spectra
are small. But this is too strong assumption. More careful analysis shows
that it is enough to have the factorization properties \C{com}:
\ba
\int \prod \{d\o (k_i) dr_i\}  <\e (k_1)\e (k_2)\cdots >|_{(r_1,r_2,...)} -
\n\\
\prod \int d\O (k_i) dr_i <\e (k_i)>|_{(r_i)} \sim 0.
\ea
It must be noted that this is the unique solution of problem since the
expansion near $\b_{1(2)}(E)$ $unavoidably$ leads to asymptotic
series with zero radii of convergence.

One can hope to avoid this problem working permanently in the energy-momentum
representation, i.e. without introduction of temperatures. Of course this
is possible in particles physics, but if $\b_{1(2)}(E)$ is not the `good'
parameter this means that all correlations between created particles are
sufficient, i.e. only the energy-momentum representation did not solve the
problem.

At the end, discussed factorization property of ${\bf D}_s$, $s>1$,
is well known Bogolyubov's condition of `shortened' description of
nonequilibrium thermodynamical systems with $s$-particle distribution
functions ${\bf D}_s$, $s>1$, expressed in terms of ${\bf D}_1$. It is the
condition for the `hydrodynamical' descriptions applicability since it
assumes that the constant $\b_1(E)=\b_2(E)$ is a `good' parameter for
description of whole system.

Considering a problem with nonzero nonequilibrium flaw it is hard to expect
that $\b_{1(2)}(E)$ is a good parameter, i.e. that the factorization
conditions are hold. Nevertheless, as was mentioned above, there is
possibility to have the {\it mean values} of correlators sufficiently small
in restricted ranges of phase space. It is the so called `kinetic' phase of
the process when the memory of initial state was disappeared, the `fast'
fluctuations was averaged over and we can consider the long-range
fluctuations only.

\section{Local equilibrium hypothesis}
\setcounter{equation}{0}

Let us return now to description of experimental situation in the high
multiplicity experiments. Having at energies of modern
accelerators thousands of particles in a final state it is a
hard problem even to count such big numbers. So, the number of particles
$n$ can not be considered as a trigger. Moreover, it seems naturally that
it is not important have
we hundred thousand of particles or hundred thousand plus one. To do first
step toward CQGP it is enough to be sure that on experiment the transition
of `hot' initial state into `cold' final one is examined. For this purpose
the ordinary calorimeters can be used \C{sys1}.

So, we must assume that the energies of created particles $\e_i \leq \e_0$,
where $\e_0$ is fixed by experiment. Then using energy conservation law
at given $\e_0$ the number of created particles is bounded from below:
$n > \sqrt{s}/\e_0 \equiv n_{min}$. With this constraint the integral
cross section
$$
\s_{\e_0}(s)=\sum_{n=n_{min}}\s_n (s)
$$
is measured. Choosing $n_{min} >> \bar{n}$, i.e. $\e_0 << \sqrt{s}/\bar{n}(s)$,
we get into high multiplicity region. There is also a theoretical possibility
to restore the quantity $\sim \s_n$ calculating the difference $\s_{\e_0}(s)-
\s_{\e_0 + \d\e_0}(s)$ \C{sys1}.

It is not necessary to measure energy of each particle to have $n_{min} >>
\bar{n}$. Indeed, let $\tilde{\e}_i$ is the energy of $i$-th group of
particles, $\tilde{\e}_1 +\tilde{\e}_2 +...+\tilde{\e}_k =\sqrt{s}$ and let
$\tilde{n}_i$ is the number of particles in the group,
$\tilde{n}_1 +\tilde{n}_2 +...+\tilde{n}_k =n$\footnote{It is assumed that the
number of calorimeter cells $K \geq k$.}. Then, if $\tilde{\e}_i <\e_0$,
$i=1,2,...,k$, we have inequality: $k > n_{min}$. Therefor, we
get into high multiplicities domain since $n \geq k$, if
$\e_0 << \sqrt{s}/\bar{n}(s)$. We can use the calorimeter demanding that the
educed in each cell energy $\tilde{\e}_i < \e_0$.

The preparation of such experiment is not hopeless task and it may be
sufficiently informative. This formulation of experiment we will put in
basis of the theory. Theoretically we should shrink the 4-dimension of
calorimeter cells up to zero since we do not know {\it ad hoc} the cells
dimension. Then the cells index $i$ is transformed
into the position of particle $r$. So we come to contradiction with quantum
uncertainty principle. This forces to use the Wigner functions formalism and
the first question which must be solved is to find a way as this formalism
can be adopted for description of our experiment (there is also interesting
ideas concerning applicability of Wigner functions in \C{man}).

	{\it 5.1. Vacuum boundary condition.}

We  start consideration from the assumption that the
temperature fluctuations are large scale. In a cell the
dimension of which is much smaller then the fluctuation scale of
temperature we can assume that the temperature is a `good' parameter.
(The `good' parameter means that the corresponding fluctuations
are Gaussian.)

Let us surround the  interaction region, i.e. the system under consideration,
by $N$ cells with known space-time position and let us
propose that we can measure the energy and momentum of groups of
in- and out-going particles in each cell. The 4-dimension of cells
can not be arbitrary small in this case because
of the quantum uncertainty principle.

To describe this situation we decompose $\d$-functions in (\r{de})
on the product of $(N+1)$ $\d$-functions:
$$
\delta^{(4)}(P-\sum^{n}_{k=1}q_k)=
\int
\prod^{N}_{\nu =1}\{dQ_{\nu}\delta (Q_{\nu}-\sum^{n_{\nu}}_{k=1}q_{k,\nu})\}
\delta^{(4)}(P-\sum^{N}_{\nu =1}Q_{\nu}),
$$
where $q_{k,\nu}$ are the momentum of $k$-th in-going particle in the
$\nu$-th cell and $Q_{\nu}$ is the total 4-momenta of $n_{\nu}$ in-going
particles in this cell, $\nu =1,2,...,N$. The same decomposition will be used for the
second $\d$-function in (\r{de}). We must take into account the multinomial
character of particles decomposition on $N$ groups. This will give the
coefficient:
$$
\frac{n!}{n_{1}!\cdots n_{N}!}\delta_{K}(n-\sum^{N}_{\nu =1}n_{\nu})
\frac{m!}{m_{1}!\cdots m_{N}!}\delta_{K}(m-\sum^{N}_{\nu =1}m_{\nu}),
$$
where $\d_{K}$ is the Kronecker's $\d$-function.

In result, the quantity
\ba
r((Q)_N,(P)_N)=
\sum_{(n.m)} \int |a_{(n,m)}|^2\times
\n\\ \times
\prod^{N}_{\nu =1}\{ \prod^{n_{\nu}}_{k=1}\frac{d\omega (q_{k,\nu})}{n_{\nu}!}
\delta^{(4)}(Q_{\nu}-\sum^{n_{\nu}}_{k=1}q_{k,\nu})
\prod^{m_{\nu}}_{k=1}\frac{d\omega(p_{k,\nu})}{m_{\nu}!}
\delta^{(4)}(P_{\nu}-\sum^{m_{\nu}}_{k=1}p_{k,\nu})\}
\l{34}
\ea
describes a probability to measure in the $\nu$-th cell the fluxes
of in-going particles with total 4-momentum  $Q_{\nu}$ and of out-going
particles with the total 4-momentum $P_{\nu}$. The sequence of this two
measurements is not fixed.

The Fourier transformation of $\d$-functions in (\ref{34}) gives:
$$
r((Q)_N,(P)_N)=\int \prod^{N}_{k=1}
\frac{d^4 \alpha_{1,\nu}}{(2\pi)^4}
\frac{d^4 \alpha_{2,\nu}}{(2\pi)^4}
e^{i\sum^{N}_{\nu =1}(Q_{\nu}\alpha_{1,\nu} +
P_{\nu}\alpha_{2,\nu})}
\R((\alpha_1)_N,(\alpha_2)_N),
$$
where
$$\R((\alpha_1 )_N,(\alpha_2 )_N )=\R(\alpha_{1,1},
\alpha_{1,2}...,\alpha_{1,N};
\alpha_{2,1},\alpha_{2,2},...,\alpha_{2,N})$$
has the form:
\ba
\R((\alpha_1)_N,(\alpha_2)_N)
=\int
\prod_{\nu =1}^{N}\{\prod^{n_{\nu}}_{k=1} \frac{d\omega(q_{k,\nu})}{n_{\nu}!}
e^{-i\alpha_{1,\nu}q_{k,\nu}}\times
\n \\ \times
\prod^{m_{\nu}}_{k=1} \frac{d\omega(p_{k,\nu})}{m_{\nu}!}
e^{-i\alpha_{2,\nu}p_{k.\nu}}\}
|a_{(n,m)}|^2.
\l{35}
\ea
Inserting (\ref{9}) into (\ref{35}) we find:
\ba
\R((\alpha_-)_N,(\alpha_+)_N)=
\exp\{ i\sum_{\nu =1}^{N} \int dx dx' [
\hat{\phi}_+ (x)D_{+-}(x-x';\alpha_{2,\nu})\hat{\phi}_- (x')-
\n\\
-\hat{\phi}_- (x)D_{-+}(x-x';\alpha_{1,\nu})\hat{\phi}_+ (x')]\}
Z(\phi_+)Z^*(\phi_-),
\l{36}
\ea
where $\p_-$ is  defined on the complex conjugate contour
$C_-:t\rightarrow t-i\e$ and $D_{+-}(x-x';\alpha)$, $D_{-+}(x-x';\alpha )$
are the positive and negative frequency correlation functions correspondingly.

We must integrate over sets $(Q)_N$ and $(P)_N$ if the
distribution of fluxes momenta over cells is not fixed. In result,
\be
r(P)=\int D^{4}\alpha_1 (P) d^{4}\alpha_2 (P)
\R((\alpha_1)_N ,(\alpha_2)_N),
\l{37}
\ee
where the differential measure
$$
D^{4}\alpha (P)=\prod^{N}_{\nu =1} \frac{d^4 \alpha_{\nu}}{(2\pi)^4}
K(P,(\alpha)_N)
$$
takes into account the energy-momentum conservation laws:
$$
K(P,(\alpha)_N)=
\int \prod^{N}_{\nu =1} d^4 Q_{\nu}
e^{i\sum^{N}_{\nu =1}\alpha_{\nu}Q_{\nu}}
\delta^{(4)}(P-\sum^{N}_{\nu =1}Q_{\nu}).
$$
The explicit integration gives that
$$
K(P,(\alpha)_N)\sim \prod^{N}_{\nu =1} \delta^{(3)}(\alpha -\alpha_{\nu}),
$$
where $\vec{\alpha}$ is the center of mass (CM) 3-vector.

To simplify the consideration let us choose the CM frame and put
$\alpha=(-i\beta ,\vec{0})$. In result,
$$
K(E,(\beta)_N)=\int^{\infty}_{0} \prod^{N}_{\nu =1} dE_{\nu}
e^{\sum^{N}_{\nu =1}\beta_{\nu}E_{\nu}}
\delta(E-\sum^{N}_{\nu =1} E_{\nu})
$$
Correspondingly, in the CM frame,
$$
r(E)=\int D \beta_1 (E) D \beta_2 (E) \R((\beta_1 )_N,(\beta_2 )_N),
$$
where
$$
D \beta (E)=\prod^{N}_{\nu=1}\frac{d \beta_{\nu}}{2\pi i}K(E,(\beta)_N)
$$
and $\R((\beta)_N)$ was defined in (\ref{36})
with $\alpha_{k,\nu}=(-i\beta_{k,\nu},\vec0),\;\;
Re\beta_{k,\nu} >0,\;\;k=1,2$.

We will calculate integrals  over $\beta_k$ using the stationary phase
method. The equations for mostly probable values of $\beta_k$:
\be
-\frac{1}{K(E,(\beta_k)_N)}\frac{\partial}{\partial \beta_{k,\nu}}
K(E,(\beta_k)_N)=
\frac{1}{R((\beta_1)_N)}
\frac{\partial}{\partial \beta_{k,\nu}}
R((\beta)_N),\;\;\;k=1,2,
\l{38}
\ee
always have the unique positive solutions $\tilde{\beta}_{k,\nu}(E)$. We
propose that the fluctuations of $\beta_{k}$ near $\tilde{\beta}_k$
are small, i.e. are Gaussian. This is the basis of the local-equilibrium
hypothesis \C {zub}. In this case $1/\tilde{\beta}_{1,\nu}$ is the temperature
in the initial state in the measurement cell $\nu$ and
$1/\tilde{\beta}_{2,\nu}$ is the temperature of the final state  in the
$\nu$-th measurement cell.

The last formulation (\ref{37}) imply that the 4-momenta $(Q)_N$ and
$(P)_N$ can not be measured. It is possible to consider another
formulation also. For instance, we can suppose that the initial set
$(Q)_N$ is fixed (measured) but $(P)_N$ is not. In this case we
will have mixed experiment: $\tilde{\beta}_{1,\nu}$ is defined by the
equation:
$$
E_{\nu}=-\frac{1}{R}
\frac{\partial}{\partial \beta_{1,\nu}}R
$$
and $\tilde{\beta}_{2,\nu}$ is defined by second equation in (\ref{38}).

Considering limit $N\rightarrow \infty$ the dimension of cells
tends to zero. In this case we are forced by quantum uncertainty
principle to propose that the 4-momenta sets $(Q)$ and
$(P)$ are not fixed. This formulation becomes pure thermodynamical:
we must assume that $(\b_1)$ and $(\b_2)$ are measurable quantities.
For instance, we can fix $(\b_1)$ and try to find $(\b_2)$ as the
function of total energy $E$ and the functional of $(\b_1)$.
In this case eqs.(\ref{38}) become the functional equations.

In the  considered microcanonical description the finiteness of
temperature does not touch the quantization mechanism. Really, one
can see from (\ref{36}) that all thermodynamical information is
confined in the operator exponent
$$
e^{\hat{N}(\phi_i^* \phi_j)}=
\prod_{\nu}\prod_{i\neq j}e^{i\int \hat{\phi}_{i} D_{ij}\hat{\phi}_{j}}
$$
the expansion of which describes the environment, and the `mechanical'
perturbations are  described by the amplitude $Z(\phi)$. This
factorization was achieved by introduction of auxiliary field $\p$ and is
independent from the choice of boundary conditions, i.e. from the choice
of considered systems environment.

	{\it 5.2. Wigner functions formalism}

We will use the Wigner functions formalism in the Carrusers-Zachariasen
formulation \C{carr}. For sake of generality the $m$ into $n$ particles
transition will be considered. This will allow to include into consideration
the heavy ion-ion collisions.

In the previous  section the generating functional $R((\beta)_N)$
was calculated by means of dividing the `measuring device'
(calorimeter) on the $N$ cells. It was assumed that the dimension
of device cells tends to zero ($N\rightarrow \infty$). Now we
will specify the cells coordinates using the Wigner's description.

Let us introduce the distribution function $F_n$ which defines the
probability to find $n$ particles with definite momentum and with
arbitrary coordinates. This probabilities (cross sections) are
usually measured in particle physics. The corresponding Fourier-transformed
generating functional can be deduced from (\ref{36}):
\ba
F(z,(\beta_+)_N ,(\beta_-)_N)=
\prod^{N}_{\nu =1}
\prod_{i\neq j}e^{\int d\omega (q)
\hat{\phi}^*_i (q)
e^{-\beta_{j,\nu}\epsilon (q)}
\hat{\phi}_j (q) z^{\nu}_{ij}(q)}\times
\n \\\times
Z(\phi_+)Z^*(\phi_-).
\l{39}
\ea
The variation of $F$ over $z^{\nu}_{ij}(q)$ generates corresponding
distribution functions. One can interpret $z^{\nu}_{ij}(q)$ as the
local activity: the logarithm of $z^{\nu}_{ij}(q)$ is conjugate to
the particles number in the cell $\nu$ with momentum $q$ for the
initial ($ij=21$) or final ($ij=12$) states. Note that
$z^{\nu}_{ij}(q)\hat{\phi}^*_i (q)\hat{\phi}_j (q)$ can be
considered as the operator of activity.

The Boltzman factor $e^{-\beta_{i,\nu} \epsilon (q)}$ can be interpreted as
the probability to find a particle with the energy $\epsilon (q)$ in the
final state ($i=2$) and in the initial state ($i=1$).
The total probability, i.e. the process of creation and
further absorption of $n$ particles, is defined by multiplication of
this  factors.

The generating functional (\ref{39}) is  normalized as follows:
\be
F(z=1,(\beta ))=R((\beta)),
\l{40}
\ee
$$
F(z=0,(\beta))=|Z(0)|^2=\R_0 (\phi_{\pm})|_{\phi_{\pm}=0}
$$
Where
$$
\R_0 (\phi_{\pm})=
Z(\phi_+)Z^*(\phi_-)
$$
is the `probability' of the vacuum into vacuum transition in presence
of auxiliary fields $\phi_{\pm}$. The one-particle distribution function
\ba
F_1 ((\beta_1)_N,(\beta_2)_N;q)
=\frac{\d}{\d z^{\nu}_{ij}(q)} F|_{z=0}=
\n \\
=\{\hat{\phi}^*_i (q)e^{-\beta^{\nu}_{i}\epsilon (q)/2}\}
\{\hat{\phi}_j (q)e^{-\beta^{\nu}_{i}\epsilon (q)/2}\}
\R_0 (\phi_{\pm})
\l{3.5}
\ea
describes the probability to find one particle in the vacuum.

So,
\ba
F_1 ((\beta_1)_N,(\beta_2)_N;q)=
\int dx dx' e^{iq(x-x')}e^{-\beta_{i,\nu}\epsilon (q)}\}
\hat{\phi}_i (x)\hat{\phi}_j (x')\R_0(\phi_{\pm})=
\n \\
=\int dr \{dy e^{iqy}e^{-\beta_{i,\nu}\epsilon (q)}\}
\hat{\phi}_i (r+y/2)\hat{\phi}_j (r-y/2)\R_0(\phi_{\pm})\}.
\l{3.6}
\ea
We introduce using this definition the one-particle Wigner function $W_1$
\C{carr}:
$$
F_1 ((\beta_1)_N,(\beta_2)_N;q)=
=\int dr W_1 ((\beta_1)_N,(\beta_2)_N;r,q).
$$
So,
$$
W_1 ((\beta_1)_N,(\beta_2)_N;r,q)=\int dy e^{iqy}e^{-\beta_{i,\nu}\e (q)}
\hat{\phi}_i (r+y/2)\hat{\phi}_j (r-y/2)\R_0(\phi_{\pm}).
$$
This distribution function describes the probability to find
in the vacuum particle with momentum $q$ at the point $r$ in the
cell $\nu$

Since the choice of the device coordinates is in our hands it is
natural to adjust the cell coordinate $\n$ to the
coordinate of measurement $r$:
$$
W_1((\beta_1)_N,(\beta_2)_N;r,q)=\int dy
e^{iqy}e^{-\beta_{i}(r)\epsilon (q)}\}
\hat{\phi}_i (r+y/2)\hat{\phi}_j (r-y/2)\R_0(\phi_{\pm}).
$$
This choice of the device coordinates lead to the following
generating functional:
\ba
F(z,\beta)=
\exp\{i\int dydr [\hat{\phi}_+ (r+y/2)D_{+-}(y;\beta_2 (r),z)
\hat{\phi}_- (r-y/2)-
\n \\
-\hat{\phi}_- (r+y/2)D_{-+}(y;\beta_1 (r),z)\hat{\phi}_+ (r-y/2)]\}
\R_0(\phi_{\pm}),
\l{3.10}
\ea
where
$$
D_{+-}(y;\beta (r),z)=-i\int d\omega (q) z_{+-}(r,q)
e^{iqy}e^{-\beta (r)\epsilon (q)},
$$
$$
D_{-+}(y;\beta (r),z)=i\int d\omega (q) z_{-+}(r,q)
e^{-iqy}e^{-\beta (r)\epsilon (q)}
$$
are the modified positive and negative correlation functions.

The inclusive, partial, distribution functions are familiar
in the particle physics. This functions describe
the distributions in presence of arbitrary number of other particles.
For instance, one-particle partial distribution function
\ba
P_{ij} (r,q;(\beta ))
=\frac{\d}{\d z_{ij}(r,q)} F(z,(\beta))|_{z=1}=
\n \\
=\frac{e^{-\beta_{i}(r)\epsilon (q)}}
{(2\pi)^3 \epsilon (q)}\int dy e^{iqy}
\hat{\phi}_i (r+y/2)\hat{\phi}_j (r-y/2)
\R(\phi_{\pm},(\beta)),
\l{3.13}
\ea
where eq.(\ref{40}) was used.

The mean multiplicity $n_{ij}(r,q)$ of particles in the infinitesimal
cell $Y$ with momentum $q$ is
$$
n_{ij}(r,q)
=\int dq \frac{\d}{\d z_{ij}(r,q)} \ln F(z,(\beta))|_{z=1}.
$$
If the interactions among fields are switched out we can find that
(omitting indexes):
$$
n(Y,q_0)=\frac{1}{e^{\beta (r)q_0}-1},\;\;q_0=\epsilon (q)>0.
$$
This is the mean multiplicity of black-body radiation.

	{\it 5.3. Closed path boundary conditions.}

The developed formalism allows to introduce more general
`closed-path' boundary conditions. Presence of external black-body
radiation flow will reorganize the differential
operator $\exp\{\hat{N}(\phi_i^* \phi_j)\}$ only and new
generating functional $\R_{cp}$ has the form:
$$
\R_{cp}(\alpha_1,\alpha_2)=e^{\hat{N}(\phi_i^* \phi_j)}
\R_0 (\phi_{\pm}).
$$
The calculation of operator $\hat{N}(\phi_i^* \phi_j)$
is strictly the same as in Sec.3. Introducing the
cells we will find that
$$
\hat{N}(\phi_i^* \phi_j)=
\int dr dy \hat{\phi}_i(r+y/2) \tilde{n}_{ij}(Y,y)\hat{\phi}_j(r-y/2),
$$
where the occupation number $\tilde{n}_{ij}$ carries the cells index
$r$:
$$
\tilde{n}_{ij}(r,y)=\int d\omega (q) e^{iqy}n_{ij}(r,q)
$$
and ($q_0 =\epsilon (q)$)
$$
n_{++} (r,q_0)=n_{--}(r,q_0)=\tilde{n}(r,(\beta_1 +\beta_2)|q_0|/2)=
\frac{1}{e^{(\beta_1 +\beta_2)(r)|q_0|/2}-1},
$$
$$
n_{+-}(r,q_0)=\Theta (q_0)(1+\tilde{n}(r,\beta_2 q_0))+
\Theta (-q_0)\tilde{n}(r,-\beta_1 q_0),
$$
$$
n_{-+}(r,q_0)=n_{+-}(r,-q_0).
$$
For simplicity the CM system was used.

Calculating $\R_0$ perturbatively we will find that
\ba
\R_{cp}(\beta)=\exp\{-iV(-i\hat{j}_+)+iV(-i\hat{j}_-)\}\times
\n \\
\exp\{i\int dr dy[\hat{j}_i (r+y/2)G_{ij}(y,(\beta (r))\hat{j}_j (r-y/2)\}
\l{4.8}
\ea
where, using the matrix notations,
\ba
iG(q,(\beta (r)))=
\left(
\matrix{
\frac{i}{q^2 -m^2 +i\e } & 0 \cr
0 & -\frac{i}{q^2 -m^2 -i\e } \cr
}\right)
+\n \\ \n \\+
2\pi \d (q^2 -m^2)
\left(
\matrix{
n(\frac{(\beta_1 +\beta_2)(r)}{2}|q_0|) &
n(\beta_1 (r)|q_0|)a_+ (\beta_1) \cr
n(\beta_2(r)|q_0|)a_- (\beta_2) &
n(\frac{(\beta_1 +\beta_2 )(r)}{2}|q_0|) \cr
}\right),
\l{4.9}
\ea
and
\be
a_{\pm}(\beta)=-e^{\beta (|q_0|\pm q_0)/2}.
\l{4.9'}
\ee
Formally this Green functions obey the standard equations in the
$y$ space:
$$
(\partial^2 -m^2)_y G_{ii}=\delta (y),
$$
$$
(\partial^2 -m^2)_y G_{ij}=0, \;\;i \neq j
$$
since $\Phi (\s_{\infty}) \neq 0$ reflects the mass-shell particles.
But the boundary conditions for this equations are not evident.

It should be underlined that in our consideration $r$ is the coordinate
of $measurement$, i.e. $r$ is as the calorimeter cells coordinate and there
is not necessity to divide the interaction region of QGP on domains (cells).
This means that $L$ must be smaller then the typical range of fluctuations
of QGP. But, on other hand, $L$ can not be arbitrary small since this will
lead to assumption of $local$ factorization property of correlators, i.e.
to absence of interactions.

So, changing $\b \rightarrow \b (r)$ we should assume that $\b_{1(2)}(r)$
and $z_{+-(-+)}(r,k)$ are constants on interval $L$. This prescription
adopts Wigner functions formalism for the case of high multiplicities. It
describes the temperature fluctuations larger then $L$ and averages the
fluctuations smaller then $L$ leading to absence, in average, of `non-
Gaussian' fluctuations.

It is the typical `calorimetric' measurement since in a dominant number of
calorimeter cells the measured mean values of energy, with exponential
accuracy, are the `good' parameters $\sim 1/\b_{2}(r,E)$. We will assume
that the dimension of calorimeter cells $L << L_{cr}$, where $L_{cr}$ is
the dimension of characteristic fluctuations at given $n$. In deep
asymptotic over $n$ we must have $L_{cr} \rightarrow \infty$. This
consideration shows that the offered experiment with calorimeter as the
measuring device of particles energies is sufficiently informative in the
high multiplicities domain.

\section{Nonstationary statistical operator}\0

One  can not expect the evident connection between the above considered
$S$-matrix (microcanonical) and Zubarev's \C{zub} approaches. The reason
is introduction into Zubarev's formalism interaction with a heat bath,
external to system under consideration. This interaction is crucial for
definition of NSL for explanation of the trend to maximal-entropy state,
starting evolution from local-equilibrium state\footnote{This condition is
not necessary in the $S$-matrix formalism since it is `dynamical' by its
nature, i.e. includes the notion of initial- and final-states as the
boundary conditions.}.

Therefore, in Zubarev's theory the local-equilibrium state was
chosen as the boundary condition. It is assumed that in the suitably
defined cells of the $system$ at a given temperature distribution
$T(\vec{x},t)=1/\beta(\vec x,t)$, where $(\vec x ,t)$ is the index of
the cell, the entropy is maximal. The corresponding nonequilibrium
statistical operator
\be
\R_z \sim e^{-\int d^{3}x \beta(\vec x,t ) T_{00}}
\l{41}
\ee
describes  evolution of a system in the time scale $t$. Here
$T_{\mu \nu}$ is the energy-momentum tensor. It is assumed that the system
`follows' to $\beta(\vec x, t)$ evolution and the local temperature
$T(\vec{x},t)$ is defined as the external parameter which is
the regulator of systems dynamics. For this purpose the special
$i\e$-prescription was introduced (it was not shown in (\r{41}) \C{zub}.
It brings the interaction with heat bath.

The KMS periodic boundary condition can not be applied for nonstationary
temperature distribution and by this reason the decomposition:
\be
\beta (\vec{x},t)=\beta_{0}+\beta_{1}(\vec{x},t)
\l{1.2}
\ee
was offered in the paper \C {paz}. Here $\beta_0$ is the constant and
the inequality
$$
\b_{0}>>|\beta_{1}(\vec{x},t)|
$$
is  assumed. Then,
\be
\R_{z}\sim e^{-\beta_{0}(H_{0}+V+B)}
\l{42}
\ee
where $H_0$ is the  free part of the  Hamiltonian, $V$ describes the
interactions and the linear over $\beta_1 /\beta_0$ term $B$ is connected
with the deviation of temperature from the `equilibrium' value $1/\beta_0$.
Presence of $B$-perturbations creates the `thermal' flows in the system to
explain increasing entropy. Considering $V$ and $B$ as the perturbations
one can calculate the observables averaging over equilibrium states, i.e.
adopting the KMS boundary condition. Using standard terminology one can
consider $V$ as the `mechanical' and $B$ as the `thermal' perturbations.

The quantization problem of operator  (\ref{42}) is connected
with definition of the space-time sequence of mechanical ($V$)
and thermal ($B$) excitations. It is necessary since the mechanical
excitations give the influence on the thermal ones and vice versa.
It was assumed in \C {paz} that $V$ and $B$ are commuting operators,
i.e. the sequence of $V$- and $B$-perturbations is not sufficient.
The corresponding generating functional has the form \C{paz}:
$$
Z(j)=\exp\{-i\int_{C_{\b}}d^4 x (V(-i\hat j(x)) +
\frac{\b_1 ({\bf x}, \tau)}{\b_0}T_{00}[-i\hat j(x)] -
$$
$$
\int^{0}_{-\infty}dt_1 \frac{\b_1 ({\bf x}, \tau +t_1)}{\b_0}
T_{00}[-i\hat j({\bf x}, x_0 t_1)])\}
Tr (e^{-\b_0 H_0}T_C e^{i\int_C d^4y j(y) \P (y)} ),
$$
where the time contour $C_{\b}$ was described in Sec.4.1 and $\tau$ is
the measurement time.

It is evident that this solution leads to the renormalization by
the interactions with the external field $\beta(\vec x, t)$ even without
interactions among fundamental fields $\P$. The source of this
renormalizations is the kinetic term in the energy-momentum tensor $T_{00}$,
i.e. follows from `thermal' interactions with external heat bath.
Note absence of this renormalizations in the $S$-matrix formalism, see, for
instance (\r{29}), where the interactions generated by $V$-perturbations
only.

In \C {bib} the operators $V$ and $B$ are  noncommuting ones and
$B$-perturbations were switched on after $V$-perturbations.
In this formulation the nondynamical renormalization are also
present but it is not unlikely that they are
canceled at the very end of calculations \C {15}.

This formulation with $\beta(\vec x, t)$ as the external field
reminds the old, firstly quantized, field theory in which matter
is quantized but fields are not. It is well known that consistent quantum
field theory requires the second quantization. Following to this
analogy, if we want to take into account consistently the reciprocal
influence of $V$- and $B$-perturbations the field $\beta(\vec{x}, t)$
must be fundamental, i.e. must be quantized (and the assumption of
paper \C {paz} becomes true). But it is evidently the wrong idea in
the canonical Gibbs formalism. So, as in the firstly quantized theory,
the theory with operator (\ref{41}) must have the restricted range of
validity \C{zub}.

\section{Conclusion}\0

In our interpretation of the  real-time  finite-temperature field theory
the statistics and the fields quantum dynamics were factorized:
statistics is fixed by the operator $\exp\{\hat{N}(\phi^*_i \phi_j\}$
and  a pure field-theoretical dynamics  is described by
$\R_0 (\phi_{\pm})=Z(\phi_+ )Z^*(\phi_-)$, where $Z(\phi_{\pm})$
is the vacuum into vacuum transition amplitude in presence of the
external (auxiliary) fields $<vac|vac>_{\phi}$. We can say that the operator
$\exp\{\hat{N}(\phi^*_i \phi_j\}$ maps the system of interacting fields
on the state with definite thermodynamical parameters. We had concentrated
our attention in this paper on the structure and origin of operator
$\exp\{\hat{N}(\phi^*_i \phi_j\}$ only and do not discuss
$\R_0 (\phi_{\pm})$. But the developed formalism allows to use following
`$S$-matrix' properties which are new for thermodynamics to define $\R_0$.

First of them is absence of Matsubara imaginary parts of time contour in
$\R_0$ by definition: the approach is pure `real-time'. This allows to
construct the formalism without referring to time asymptotic properties of
correlation (Green) functions, and introduce the temperature description
without using a notion of grand canonical ensemble constructing the
environment of the system, i.e. the measuring device, `by hand'.

Moreover, discussed factorization property have important consequence which
would allow to calculate expectation values with high accuracy. Let us
consider the theoretical problem of the $\R_0 (\phi_{\pm})$ calculation.
To define the functional measure the ortho-normalizability (i.e. the
unitarity) condition may be used. It leads to following representation
\C{untr}:
\be
\R_0(\p)=e^{-i\hat{K}(j,e)}\int DM(\P) e^{-U(\P ,e)} e^{\int dx (v'(\P)+j)\p},
\l{43}\ee
where the expansion over operator
$$
\hat{K}(j,e)=2Re\int dx \f{\d}{\d j(x)}\f{\d}{\d e(x)}
$$
generates perturbation series and
$$
U(\P ,e)=V(\P +e)-V(\P-e)-2Re\int dx e v'(\P)
$$
weights quantum fluctuations. The most important term in (\r{43}) is the
measure
$$
DM(\P)=\prod_x d\P(x) \d (\pa_{\mu}^2 \P +m^2\P +v'(\P) -j)
$$
where $v'(\P) \equiv \d V(\P) /\d \P (x)$. So, solving the equation
\be
\pa_{\mu}^2 \P +m^2\P +v'(\P) =j
\l{44}\ee
we will find $all$ contributions\footnote{This means that the unitarity
condition is necessary and sufficient for definition of path integral
measure for $R_0 (\phi_{\pm})$ \C{yad}}.

At the very end of calculations one must put $e =j=0$. Therefore, eq.(\r{44})
can be solved expending it over $j$. This shows that (\r{43}) restores at
$j=0$ the usual stationary phase method. Indeed, it can be verified
that (\r{43}) gives usual perturbation theory \C{untr}.

But the eq.(\r{44}) gives much more possibilities. Note that l.h.s.
of this equation is sum of known classical forces and the r.h.s. is the
quantum force $j$. Eq.(\r{44}) establish the local equilibrium between this
forces. This solves the old standing problem of quantization with
constraints: it can be done by field transformations in path integrals
since the eq.(\r{44}) shows the way as $j$ must be transformed when the
l.h.s. is transformed. Presence of derivatives in (\r{44}) shows that the
quantum force must be transformed in the tangent space of fields\footnote
{This explains why the ordinary transformation of path integral is
impossible, gives wrong result \C{mari}.}.

The r.h.s. of eq.(\r{44}) may contain also an additional force to describe
the external influence on the system of interacting fields. This force
was omitted in eq.(\r{44}) assuming that a process of particles creation
(and absorption) switched on adiabatically.

As was mentioned above the action of operator $e^{-\hat{N}(\b,z;\p)}$ on
$R_0 (\p)$ maps interacting fields system on measurable states. Let us
consider what this gives. Result of action has the form:
$$
\R(\b,z)=e^{-i\hat{K}(j,e)}\int DM(\P) e^{-U(\P ,e)} e^{-N(\b,z.;\P)},
$$
where $N=N_1 +N_2$ and
\be
N_{1(2)}(\b,z;\P)=\int dr d\o(k)
e^{-\b_{1(2)}(r)\e (k)}z_{+-(-+)}(k,r) |\Ga(k,\P)|^2.
\l{45}\ee
Here $r$ is considered as the $index$ of calorimeter cell. This
formulae needs more careful explanation. Instead of (\r{45}) we
must consider
\ba
N_{1(2)}(\b,z;\P)=\int dr d\o(k) e^{-\b_{1(2)}(r)\e (k)}z_{+-(-+)}(k,r)
\n\\
\int  dq \d_L (q) \Ga(k+q,\P) \Ga^* (k-q,\P).
\ea
where $L$ is the scale where $\b_{1(2)}(r)$ and $z_{+-(-+)}(k,r)$ can be
considered as the constants ($L$ is the dimension of
calorimeter cell). If $L \rightarrow \infty$ then $\d_L (q)$ can be
changed on usual $\d$-function $\d (q)$ and, therefore, in this limit we
will have (\r{45}). We had considered this limit supposing that the
measurement is not in contradiction with quantum uncertainty principle.

So, deriving $N_{1(2)}(\b,z;\P)$ there was used the condition that $r$ is
the coordinate of size $L$ cell. With this condition
\be
\Ga (k, \P) =\int dx e^{ikx} (\pa_{\mu}^2 +m^2)\P
\l{46}\ee
can be considered as the order parameter. Indeed, $\Ga (k,\P)$ is the element
of actions symmetry group since it is linear over field $\P$ and the
generating functional $R(\b,z)$ is trivial if $<|\Ga (k,\P)|^2>=0$. In this
case there is not creation of particles, i.e. there is not measurable
asymptotic states (fields).

Indeed, it can bee shown \C{wig2} that $all$ quantum corrections to solitons
contribution in the (1+1)-dimensional sin-Gordon model
equal to zero. This is in accordance with result of \C{das} and
with factorizability of solitons $S$-matrix. Then it is easily seen
computing integral in (\r{46}) by parts that $\Ga (k, \P_s )=0$, where
$\P_s$ is the soliton solution.
This result shows that hidden symmetry of sin-Gordon model can not
be broken and corresponding (polynomial) integrals of motion are conserved.
The application of this idea for non-Abelian field theory should be fruitful.

\newpage

{\Large \bf Acknowledgment}
\vspace{0.2in}

I would  like to thank my colleagues from the Inst. of Phys.
(Tbilisi), Inst. of  Math. (Tbilisi), Inst. of Nucl. Phys.
(St.-Petersburg), Joint Inst. of Nucl. Res. (Dubna), Inst. of Exp.
and Theor. Phys. (Moscow), Inst. of Theor. Phys. (Kiev) and
DAMPT (Cambridge), and especially I.Paziashvili and
T.Bibilashvili, for
fruitful discussions which gave me the chance to extracting
the questions that were necessary to show up in this paper.
I would like also to thank A.Sissakian for fruitful collaboration in
the problem of high-multiplicity hadron reactions. It must be
underlined the stimulating interest to this work by V.Kadyshevski. The
experimental side of the problem I learn from T.Lomtadze and T.Grigalashvili.
I want to mention also the important information about Wigner functions
applicability in the modern solid state physics from M.Tomak. The work
was supported in part by the  U.S. National Science Foundation and
was granted in part by the Georgian Academy of Sciences.

\end{document}